\documentclass[twocolumn,secnumarabic,amssymb, nobibnotes, aps, prd]{revtex4-2}

\usepackage{graphicx, bm}
\begin{document}

\title{Splitting of energy levels of Spin-vortex Induced Loop Currents by feeding external currents}%

\author{Hikaru Wakaura}%
\email[Quantscape: ]{
hikaruwakaura@gmail.com}
\affiliation{QuantScape Inc. QuantScape Inc., 4-11-18, Manshon-Shimizudai, Meguro, Tokyo, 153-0064, Japan}

\author{Takao Tomono}

\affiliation{ Digital Innovation Div. TOPPAN Inc. TOPPAN, 1-5-1, Taito, Taito, Tokyo, 110-8560, Japan}
\email[TOPPAN: ]{takao.tomono@ieee.org}
\date{December 2021}%

\begin{abstract}
The spin-vortex-induced loop current (SVILC) is a nano-sized loop current predicted to exist in the CuO$_2$ plane in the bulk of hole-doped cuprate superconductors. It is a persistent loop current protected by the topological winding number associated with the wave function. It exists around a spin-vortex created by the itinerant electrons with a doped hole at its center. The direction of each SVILC can be either clockwise (winding number is $-1$) or counterclockwise (winding number is $+1$); and the winding number with no current (winding number is zero) is forbidden by the single-valued requirement of the wave function with respect to the electron coordinates.
Recently, it has been demonstrated, theoretically, that this degree-of-freedom can be used for qubits. Coupling of neighboring qubits by external current feeding is confirmed to be realizable.
This means that nano-sized couplers of SVILC qubits using feeding external currents are realizable. The size of couplers of SVILC qubits can be conparable or smaller than that of trapped ion qubits. Couper size of SVILC qubits is decided by the range of spin vortices in CuO$_2$ plane and current distribution, thus, this is tunable by feeding external current and substituting Cu atoms in barrier atoms. That of trapped ion qubits is limited by the distance that conbined vibration occurs or laser range with respect to the coordinates. 
In the present work, We demonstrated splitting energy levels by exernal feeding current of three qubit system of SVILC qubits. This means that nano-sized qubit differentiator can be realized, and noise by static magnetic field can be cut off, and this may enable the realizing fully-fault torelant quantum computers by SVILC qubits. Moreover, the possiblity of downscaling of them is shown.
\end{abstract}

\maketitle
\tableofcontents

\section{Introduction}\label{1}
The number of qubits in quantum computers just has reached 79 recently, where the memory space of quantum computers eclipse that of classical supercomputers. Since the feasibility of quantum error correction is demonstrated, experimentally \cite{Kelly2015}, the next step is to equip with fault tolerance. This will eventually lead to a quantum computer with a large number (say,100) logical qubits \cite{Gambetta2017}. 

In order to realize such a fault tolerant quantum computer, even the second most promising qubits today, superconducting qubits using Josephson junctions of the BCS superconductors, may not be good enough \cite{PhysRevLett.105.100502}. 
The reasons are following; 1) the size of a qubit with including coupler may be too large to accommodate a large number of qubits. The 
typical size of a supercondcting qubit with including coupler is in the order of millimeters; thus, the size of a 100 logical qubit system become in the order of meters. 2) the operation temperature may be too low. The operation temperature for the superconducting qubit is in the order of mK \cite{Linke28032017}. Maintaining a large number of qubits intact with keeping such low temperature may be too difficult and too expensive \cite{Awschalom1174}. 
3) the coherence time is too short. The longest coherent time achieved for the superconducting system is about 100 $\mu$s. 
This is enough to accommodate error corrections for a small number of qubits; however, for a large number of qubit system, a much longer coherence time may be required.

The most promising is trapped ions by electric field, realized 79-qubit system, have a potential to realize this, are not enough too. The reasons are following; 1) the size of systems can't be minimized because each qubit is differenciated by the position in the system and the distance between other ions\cite{2009arXiv0904.2599L}. 2) Qubit operation can be realized by only irradiating lasers and shortening the distance between two ions. Lasers can control the qubit states enough only in extreamely low temperature because of thermal fluctuation of energy difference\cite{brown_co-designing_2016}.

 Recently, a new qubit has been proposed that may overcome the above mentioned difficulties.
 It utilizes nano-sized persistent loop currents predicted to exist in the cuprate superconductor (Wakaura et al. Ref.~\cite{Wakaura201655}). 
 Superconductivity in the cuprate (the cuprate superconductivity) is believed to be different from the conventional one explained by the BCS theory (the BCS superconductors). 
 It shows marked differences in many properties compared to those found in the BCS superconductors.  For example, the normal state from which the superconducting state emerges is not a band metal.  A theory for the cuprate superconductivity has been proposed where it is explained from the view point of the appearance of a new current generation mechanism that occurs when holes are doped in the Mott insulator. In the new theory, the persistent loop current plays the role of the current element, and a macroscopic current is generated as a collection of them \cite{HKoizumi2013}\cite{HKoizumi2014}\cite{HKoizumi2015B}.
  
 The  loop current is called, the spin-vortex induced loop current (SVILC), and it is predicted to exist in the CuO$_2$ plane of the bulk of hole-doped cuprate superconductors. It is induced by the spin-vortex created by the itinerant electrons with a doped hole at its center. It is protected by the topological winding number associated with the wave function. The direction of each SVILC can be either clockwise (winding number $-1$) or counterclockwise (winding number $+1$); and the currentless winding number zero is forbidden by the single-valued requirement of the wave function with respect to the electron coordinates.
 
 Although the presence of the SVILC is not verified, yet, there is evidence that such loop currents exist: 1) the existence of loop currents is inferred by neutron scattering measurement  \cite{Mangin-Thro2015}; 2) the magnetic excitation spectrum calculated by assuming the existence of spin-vortices agrees with that obtained by the neutrons scattering experiment, \cite{Hidekata2011}, indicating the presence of spin-vortices is very plausible; 3) the polar Kerr effect measurement \cite{PhysRevLett.100.127002} and the enhanced Nernst effect measurement \cite{Nernst2} suggest the presence of loop currents. 
 
Assuming the existence of the SVILC, we have demonstrated, theoretically, that SVILCs can be used as qubits \cite{Wakaura201655}. We list expected properties of the SVILC qubits, below.

\begin{enumerate}
\item All qubits can be differentiated in a controlled manner by modifying the environment of them. The controlled modification can be achieved by applying external magnetic field and feeding external currents as we will show in this work. The controlled modification by external electric field is also can be achieved.

\item The qubit operation can be achieved by irradiating an electromagnetic field with the frequency that corresponds to the energy difference between the two qubit states. This can also be achieved by varying the magnitude of feeding external current to cross the energy levels of given qubit for the time correspond to qubit operation by Landau-Zenner effect\cite{PhysRevB.70.201304}.

\item Our previous calculation indicates that the gate-operation time is in the nanosecond order when an electromagnetic field with electric field intensity $10^5$ V/m is used \cite{Wakaura201655}.

\item Coupling between qubits can be turned-on and off in a controlled manner. The coupling is turned-off by placing qubits at a distance; the separation distance may be shortened by using barrier atoms (substituted atoms for Cu's) between the qubits. The coupling is turned-on by feeding external currents in the region between the two target qubits\cite{Wakaura_2017}.

\item The size of each qubit is about 10 nm$^2$. The size of the qubit-coupler using the external current feeding is also in the nanometer scale.

\item The stabilization temperature for the SVILCs corresponds to the superconducting transition temperature $T_c$ for the cuprates \cite{HKoizumi2015B} which is above the liquid nitrogen temperature. Thus, the qubit operation at temperatures above the liquid nitrogen temperature might be possible. 

\item Readout process will be performed by measuring the magnetic field produced by SVILCs after turning-off the applied magnetic field and feeding external currents. It is also possible to use the response currents to the feeding external currents \cite{Morisaki2017}. 

\item The SVILC is protected by the topological winding number, thus, it is expected to be robust against external perturbations.
Besides, it does not require the Cooper pair formation, thus, it is free from the relaxation caused by unpaired electrons that is believed to be
the major cause of limiting coherent time for the superconducting qubits using the Josephson junctions  \cite{Aumentado2004,deVisser2011,Maisi2013,Levenson-Falk2014,Gustavsson2016}.

\end{enumerate}

The purpose of the present work is to demonstrate, theoretically, that differenciating the environments of three qubits placed at long distances and uncoupled can be performed by feeding external currents coincide with coupling of qubits by also external currents. 
We also consider the substitution effect of the Cu atoms by some other atoms (we call them, `barrier atoms').
This method we propose in this paper may omit magnetic field and its generators from quantum computers using SVILCs as qubits. This also may realize the minimization of whole system of qubits.

The organization of this work is following: we explain the calculation method of the quantum states with SVILCs in Section \ref{2}. We consider the spin-vortex quartet (SVQ), a unit of four spin-vortices and their SVQ qubits, in Section \ref{3}. We construct 8 orthogonal states of three qubit system of SVQ qubits and calculate the electric transition dipole moments between them. We also perform coupling of two and three qubits. In Section \ref{4}, we perform the differenciation of the environments of qubits by external currents. We show the possibility of differentiation of each qubit coincide with coupling of qubits.
In Section \ref{6}, we conclude the present work. 
\section{Calculation of States with SVILCs}
\label{2}
In this section, we explain how to calculate the spin-vortex-induced loop current (SVILC) states. The method and underlying assumptions are already explained in our previous works \cite{Wakaura201655,HKoizumi2013,HKoizumi2014,HKoizumi2015B,Morisaki2017}. 

The Hamiltonian we will use is a Hartree-Fock version given by
\begin{eqnarray}
 &&H^{HF}_{\rm EHFS}\!=\! -t \sum_{ \langle i, j \rangle_1, \sigma} \left( c_{i \sigma}^{\dagger}c_{j \sigma} + c_{j \sigma}^{\dagger}c_{i \sigma} \right)
 \nonumber
 \\
&+&U\sum_j \Big[({{n_j} \over 2}\!-\!S_j^z)c_{j \uparrow}^{\dagger}c_{j \uparrow} \!+\!({{n_j} \over 2}+ S_j^z)c_{j \downarrow}^{\dagger}c_{j \downarrow} \nonumber \\ 
&\!-\!&(S_j^x \!-\!i S_j^y)c_{j \uparrow}^{\dagger}c_{j \downarrow}\!-\!(S_j^x \!+\!i S_j^y)c_{j \downarrow}^{\dagger}c_{j \uparrow}
\Big]
\label{HF}
 \end{eqnarray}
 where $c_{j\sigma}^{\dagger}$ and $c_{j\sigma}$ are the creation and annihilation operators of electron at the $j$th site with the z-axis projection of electron spin $\sigma$, respectively; we define $i$ and $j$ as the sites of Cu atoms,
by considering a single CuO$_2$ plane in the bulk, and ignoring the oxygen atoms between Cu atoms; the sum of $\langle i, j \rangle_1$ in the first term indicates that the sum is taken over the nearest neighbor pairs;
  \begin{eqnarray}
n_j=\sum_{\sigma} \langle c_{j \sigma}^{\dagger}c_{j \sigma} \rangle
\end{eqnarray}
is the number operator of electrons at the $j$th site, and $\bm{S}_j=(S_j^x,S_j^y,S_j^z)$ is the electron spin at the $j$th site given by 
\begin{eqnarray}
S_j^x &=& { 1 \over 2} \langle c_{j \uparrow}^{\dagger}c_{j \downarrow} +c_{j \downarrow}^{\dagger}c_{j \uparrow}
\rangle =S_j \cos \xi_j \sin \zeta_j
\nonumber
\\
S_j^y &=&- { i \over 2} \langle c_{j \uparrow}^{\dagger}c_{j \downarrow} -c_{j \downarrow}^{\dagger}c_{j \uparrow}
\rangle=S_j \sin \xi_j \sin \zeta_j
\nonumber
\\
S_j^z &=& { 1\over 2} \langle c_{j \uparrow}^{\dagger}c_{j \uparrow} -c_{j \downarrow}^{\dagger}c_{j \downarrow}
\rangle=S_j \cos \zeta_j
\label{S}
\end{eqnarray}
where $\xi$ and $\zeta$ are the azimuth and polar angles, respectively; $\langle\hat{O}\rangle$ denotes the expectation value of operator $\hat{O}$.
The parameters $n_j$, $S_j^x$, $S_j^y$ and $S_j^z$ are obtained self-consistently. 

As for materials parameter values, we use the lattice constant $a=0.4$ nm, the transfer integral $t=130$ meV, and the on-site Coulomb repulsion $U=8t$ \cite{Yamaji2011}. We also consider only the case where $\zeta_j=\pi/2$ at all sites. In this case the spin is polarized in the CuO$_2$ plane as is observed in the parent compound of the cuprate \cite{NeutronRev}. There must be a mechanism that stabilizes such spin polarization; however, we take it as a phenomenological condition without identifying it and including a responsible interaction in this work.
 
From an energy minimization requirement, we obtain single-particle wave functions,
 \begin{eqnarray}
 \mid \tilde{\gamma} \rangle& =&\sum_{j} [e^{-i{{\xi_j} \over 2}}D_{j \uparrow}^{\gamma} {c}^{\dagger}_{j \uparrow}\!+\!e^{i{{\xi_j} \over 2}}D_{j \downarrow}^{\gamma} {c}^{\dagger}_{j \downarrow}] \mid {\rm vac} \rangle,
\label{c2wave}
 \end{eqnarray}
 where $D_{j \uparrow}^{\gamma}$, $D_{j \downarrow}^{\gamma}$ and $\xi_j$ are obtained numerical parameters. 
 The many-electron wave function is usually constructed as a Slater determinant of them. 
 However, the above single-particle wave functions are multi-valued with respect to electron coordinates due to the presence of spin-vortices; the obtained many-electron wave function is also multi-valued. This contradicts the requirement that the total wave function should be a single-valued wave function.

Let us explain howe the multi-valuedness of the wave function $\mid \tilde{\gamma} \rangle$ arises, and
a way to remedy it.  A spin-vortex in the CuO$_2$ plane is described by a non-zero winding number of $\xi$ for some loop $C_{\ell}$ defined by
\begin{eqnarray}
w_{\ell}[\xi]={ 1 \over {2\pi}} \sum_{i=1}^{N_{\ell}} ( \xi_{C_{\ell}(i+1)} -\xi_{C_{\ell}(i)}),
\end{eqnarray} 
where $N_{\ell}$ is the number of lattice points for $C_{\ell}$; $C_{\ell}(i)$ denotes the $i$th lattice point in $C_{\ell}$ with the boundary condition $C_{\ell}(N_{\ell}+1)=C_{\ell}(1)$.

When the phase factor $e^{\pm {i \over 2} \xi_{C_{\ell}(1)}}$ is evaluated starting from $C_{\ell}(1)$ by integrating $\nabla \xi \approx (\xi_{C_{\ell}(i+1)} -\xi_{C_{\ell}(i)})$ along closed path $C_{\ell}$, it becomes
\begin{eqnarray}
e^{\pm {i \over 2} (\xi_{C_{\ell}(1)}+2\pi w_{\ell}[\xi])}=(-1)^{w_{\ell}[\xi]} e^{\pm {i \over 2} \xi_{C_{\ell}(1)}}
\end{eqnarray}
Then, if ${w_{\ell}[\xi]}$ is odd, sign-change occurs. This means that the wave function $\mid \tilde{\gamma} \rangle$ in Eq.~(\ref{c2wave}) is multi-valued.

We remedy the multi-valuedness by constructing the single-valued single-particle wave function $\mid \gamma \rangle$ as
 \begin{eqnarray}
 \mid \gamma \rangle& =&\sum_{j} e^{\!-\!i { {\chi_j } \over 2}} [e^{-i{{\xi_j} \over 2}}D_{j \uparrow}^{\gamma} {c}^{\dagger}_{j \uparrow}\!+\!e^{i{{\xi_j} \over 2}}D_{j \downarrow}^{\gamma} {c}^{\dagger}_{j \downarrow}] \mid {\rm vac} \rangle,
\label{Cwave}
 \end{eqnarray}
 by including the phase factor $ e^{\!-\!i { {\chi_j } \over 2}}$.
 The single-valued many-electron wave function is given as the Slater determinant of them.
 
The angular variable $\chi$ is determined in the following manner;
first, note that the single-valued requirement is satisfied if the following condition is satisfied by $\chi$; 
\begin{equation}
w_{\ell}[\chi] + w_{\ell}[\xi] = \mbox{even for all loops $C_{\ell}$}
 \label{constraint}
\end{equation}

The angular variable $\chi$ is determined by minimizing the total energy with taking into account the above condition.
This is achieved by minimizing the following functional, 
\begin{eqnarray}
F[\nabla \chi ]=E[\nabla \chi]+\sum_{\ell=1}^{N_{\rm loop}} { {\lambda_{\ell}}}\left( \oint_{C_\ell} \nabla \chi \cdot d {\bf r}-2 \pi \bar{w}_{\ell} \right),
\label{functional0}
\end{eqnarray}
where $E[\nabla \chi]$ is the total energy depends on $\nabla \chi$; the second term is the one arising from the constraint with $\lambda_{\ell}$ being the Lagrange multiplier; $\bar{w}_\ell$ is the winding number of $\chi$ along a loop $C_{\ell}$ that satisfies the constraint in Eq.~(\ref{constraint}); $N_{\rm loop}$ is the number of independent loops, where any loop in the system can be constructed by combining the $N_{\rm loop}$ independent loops. 

A set of values for $\bar{w}_\ell$ in Eq.~(\ref{functional0}) specifies a particular current distribution; in other words, by changing the values of $\bar{w}_\ell$ for the independent loops, solutions with different current distributions are obtained. In the following, we obtain states with different current patterns by changing only $\nabla \chi$, and fixing values of $D_{j \sigma}^{\gamma}$ and $\xi_j$. From the stationary condition of $F[\nabla \chi]$, $\nabla \chi$ is obtained as the solution of
\begin{eqnarray}
0={{\delta F[\nabla \chi]} \over {\delta \nabla \chi}}={{\delta E[\nabla \chi]} \over {\delta \nabla \chi}}+\sum_{\ell=1}^{N_{\rm loop}} { {\lambda_{\ell}}} {{\delta } \over {\delta \nabla \chi}} \oint_{C_\ell} \nabla \chi \cdot d {\bf r}
\label{lagchi}
\end{eqnarray}
with the constraint,
\begin{eqnarray}
\oint_{C_\ell} \nabla \chi \cdot d {\bf r}-2 \pi \bar{w}_{\ell}=0, \;  \ell=1,\cdots,N_{\rm loop}.
\label{winding}
\end{eqnarray}
From Eq.~(\ref{lagchi}), the current density is given by
\begin{eqnarray}
{\bf j}=-{{2e} \over {\hbar}} \sum_{\ell=1}^{N_{\rm loop}} { {\lambda_{\ell}}}{{\delta} \over {\delta\nabla \chi}} \oint_{C_\ell} \nabla \chi \cdot d {\bf r}.
\label{currentK}
\end{eqnarray}
This clearly indicates that the current is generated by the phase factor $e^{-{i \over 2} \chi_j }$ in a non-perturbative way. It is given as a sum of loop currents called, `spin-vortex-induced loop currents (SVILCs)'.  

The feeding external current contribution is included by adding extra loops in Eq.~(\ref{functional0}). For the detail of the method, consult Refs.~\cite{HKoizumi2014,Morisaki2017}.

\section{Three-qubit system}\label{3}
\begin{figure*}
 \begin{center}
\includegraphics[scale=0.40]{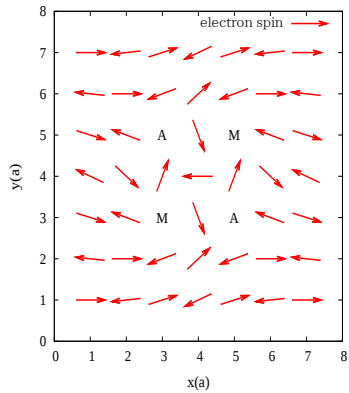}
 \end{center}
 \caption{
Spin structure of SVQ. An arrow on each lattice point indicates the electron spin whose direction is specified by $\xi_j$, where $j$ is the site index. `M' and `A' denote spin-vortex with $w_l[\xi]=1$ and $w_l[\xi]=-1$, respectively, where $\ell$ denotes the loop of 8 sites around each center. The unit of distance between two sites is 0.4$(nm)$.
 }
 \label{SVQxi}
\end{figure*}

We discuss three-qubit system of SVILC qubits here. Spin-Vortex Quartet (SVQ) is the smallest unit of spin-vortices that SVILC flows around here as shown in Fig.~\ref{SVQxi}. 

The spin-vortex quartet (SVQ) we consider is shown in Fig.~\ref{SVQxi}. It is composed of four SVILCs. The sum of the winding numbers for the spin-vortices is zero, and the nearby spin-vortices have the opposite winding numbers. This arrangement of the spin-vortices is the minimal energy one. 

\begin{figure*}
\begin{center}
\includegraphics[scale=0.3]{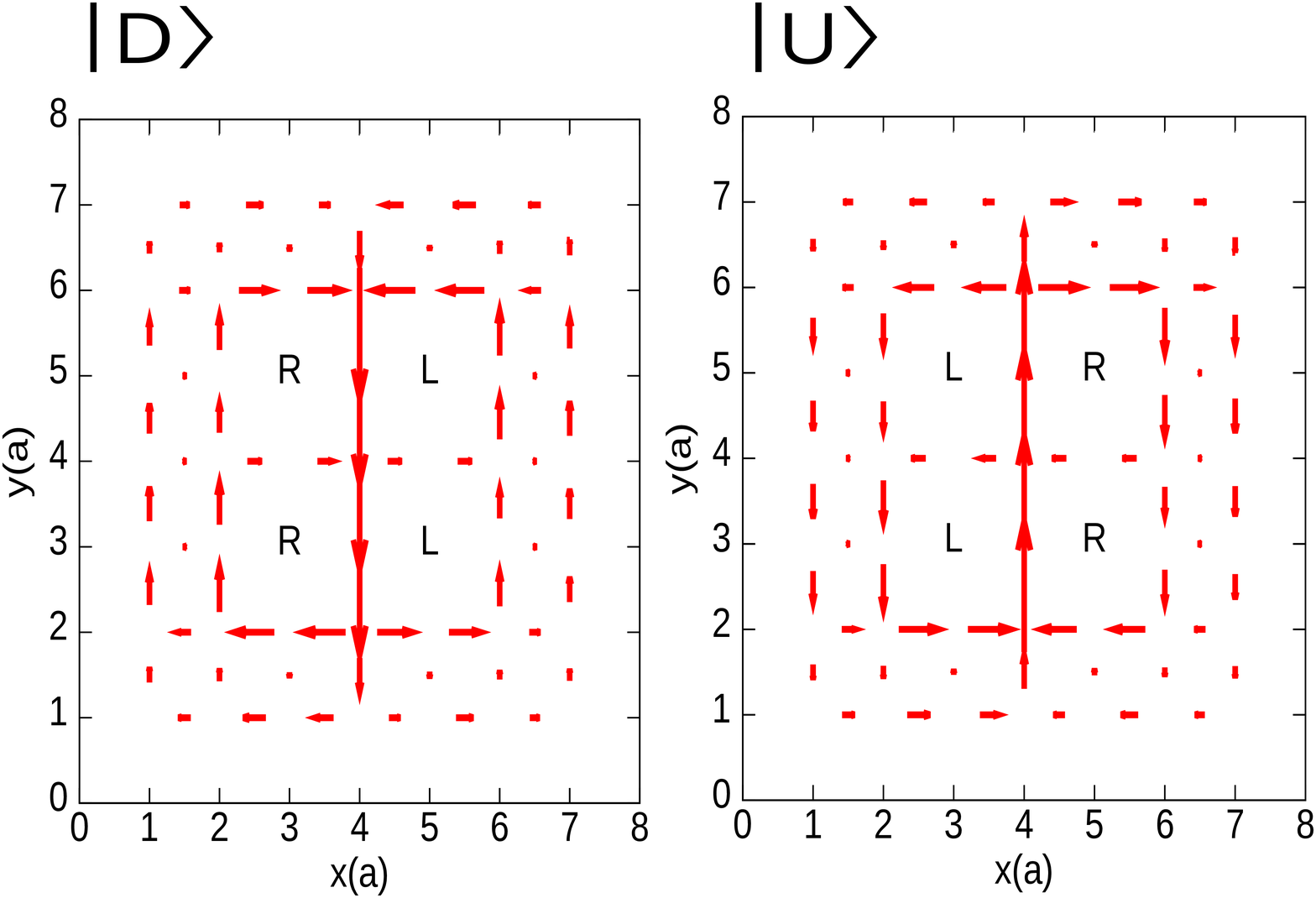}
\end{center}
\caption{
 2 current patterns for DCQ qubits. These 2 states are derived by applying a magnetic field in the $z$ direction given by $B=0.178x^2+6.0x+0.07 y^2 + 6.0 y$ T (the unit of $x$ and $y$ is $a$, where $a=0.4(nm)$ is the lattice constant of the CuO$_2$ plane).
 SVILCs are depicted by arrows. `L' and `R' denote the centers of SVILCs with $w_l[\chi]= 1$ and $w_l[\chi]=-1$, respectively, where $w_l[\chi]$ is the winding number of $\chi$ around each center. 
}\label{DCQ}
\end{figure*}

The winding number of $\chi$ for each spin-vortex is either $+1$ or $-1$, thus, totally 16 different SVILC states are possible from a single SVQ. Especially, $y$ component of transition dipole moment between upward directed current state and downward directed current state is significantly large, thus, we use these two state as two states of single qubit and define as U state and D state, respectively as shown in Fig.~\ref{DCQ}. We call this Dipole-Current-Qubit (DCQ). 

The many-electron wave functions constructed using $\{ | \gamma \rangle \}$ are $\{ \mid \tilde{\Phi}_{\rm a} \rangle \}$, where `a' indicates the current pattern specified by the winding numbers, and the angular variable $\chi$ is obtained separately depending on the current pattern. From these, we construct $\{\mid \Phi_{\rm a} \rangle \}$ by diagonalizing the matrix for the operator $H_{\rm B}$ whose $({\rm a},{\rm b})$ element is given by $\langle \tilde{\Phi}_{\rm a}\mid H_{\rm B} \mid \tilde{\Phi}_{\rm b} \rangle$. The energy of the $\mid \Phi_{\rm a} \rangle$ state is calculated as $ \langle \Phi_{\rm a} \mid H^{HF}_{\rm EHFS}+H_{\rm B} \mid \Phi_{\rm a} \rangle$ \cite{Wakaura201655}. 
We use the orthogonal basis $\{\mid \Phi_{\rm a} \rangle \}$ as qubit states.

\begin{figure*}
 \begin{center}
\includegraphics[scale=0.25]{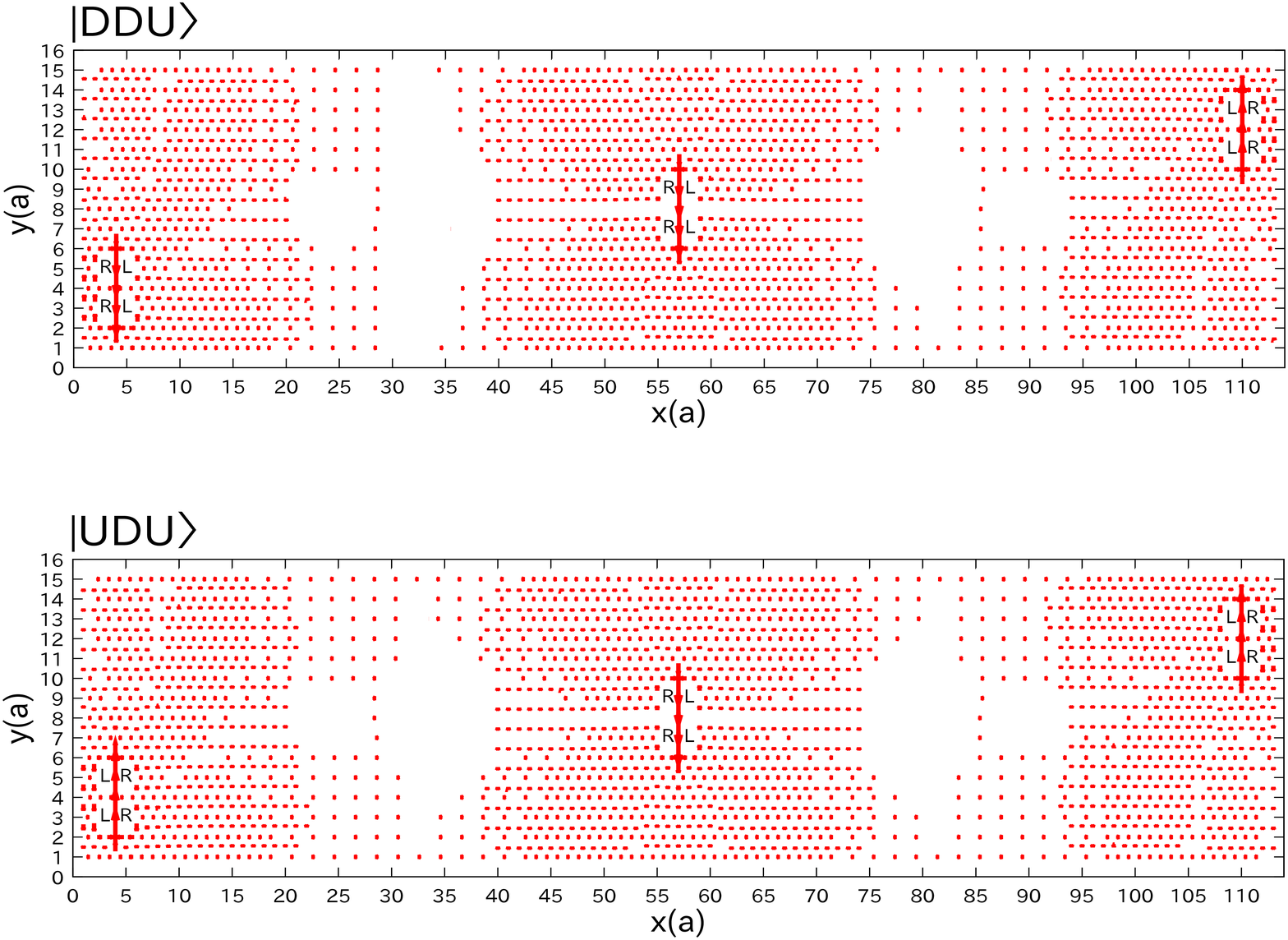}
\newline
 \end{center}
 \caption{
The current distribution of bottom two states of the three DCQ qubit system. One lattice distance corresponds to the current of the magnitude $1/3$ in the units of $2et/\hbar$. The two states are indicated as $\mid {\rm DDU} \rangle$ and $\mid {\rm UDU} \rangle$, respectively. The centers of the three DCQs are $(x,y)=(4,4)$,$(57,8)$ and $(110,12)$. 
 }
 \label{3qubitDCQb2}
\end{figure*}

Three-qubit system is three DCQs put with respect to the center of the system in 112a$\times$14a sized CuO$_2$ plane. Bottom two states of 8 quantum state of three-qubit system are shown in Fig.~\ref{3qubitDCQb2}. Three DCQs in this three-qubit system are enough separated and uncoupled. Cu atoms at $(30,1)$,$(30,2)$,$\cdots$,$(30,15)$,$(32,1)$,$(32,2)$,$\cdots$,$(32,15)$,$(82,1)$, $(82,2)$,$\cdots$,$(82,15)$,$(84,1)$,$(84,2)$,$\cdots$,$(84,15)$ are substituted by barrier atoms.

We apply a quadratic inhomogeneous magnetic field in the direction perpendicular to the CuO$_2$ plane to remove all degeneracies of the SVILC states, and construct orthogonal states that can be used as qubit states. We call them SVQ states or SVQ qubit states.

The inhomogeneous magnetic field we apply is
\begin{equation}
B=4.005x^2+540.0x+1.575y^2+135.0y \quad \mbox{(T)}\label{QDB}
\end{equation}
 where $x$ and $y$ are given in the units of lattice constant $a$.

The interaction Hamiltonian between the electric current and the applied magnetic field $\nabla \times {\bf A}^{\rm em}=(0, 0, B)$ is given by
\begin{eqnarray}
 {H}_{\rm B}=-\sum_{\langle k, j \rangle_1} \int_{\bm{r}_j}^{\bm{r}_k}
 {\bf A}^{\rm em} (\bm{r})\cdot d\bm{r}\hat{j}_{k \leftarrow j},
 \label{int}
\end{eqnarray}
where $\hat{j}_{k \leftarrow j}$ is the current operator for the current from the site $j$ to its nearest neighbor site $k$ given by
\begin{eqnarray}
\hat{j}_{k \leftarrow j}=ie{\hbar}^{-1}t\sum_{\sigma}(c^{\dagger}_{k \sigma}c_{j \sigma}-c^{\dagger}_{j \sigma}c_{k \sigma})
\end{eqnarray}
The coordinates for $j$ and $k$ are denoted as ${\bm{r}_j}$ and ${\bm{r}_k}$, respectively.

 Feeding external currents split energy levels of qubit states and expand current regions of two DCQs and couple in case they are neighboring. As shown in Fig.~\ref{DCQ3qiqsepex}, the magnitude of external currents for splitting energy levels of left DCQ is defined as $J_{\rm ex}^{\rm 1}$; they are feeding external currents at $(2,1)$ and $(6,15)$ and draining external currents at $(6,1)$ and $(2,15)$. The magnitude of external currents for splitting energy levels of center DCQ is defined as $J_{\rm ex}^{\rm 2}$; they are feeding external currents at $(55,1)$ and $(59,15)$ and draining external currents at $(59,1)$ and $(55,15)$. The magnitude of external currents for splitting energy levels of light DCQ is defined as $J_{\rm ex}^{\rm 3}$; they are feeding external currents at $(108,1)$ and $(112,15)$ and draining external currents at $(112,1)$ and $(108,15)$. The magnitude of external currents for coupling left and center DCQs is ${\rm J}_{\rm ex}^{\rm 4}$; they are feeding external currents at $(7,1)$ and $(31,15)$ and draining external currents at $(31,1)$ and $(41,15)$. The magnitude of external current for coupling center and right DCQs is ${\rm J}_{\rm ex}^{\rm 5}$; they are feeding external currents at $(83,1)$ and $(105,15)$ and draining external currents at $(71,1)$ and $(83,15)$, respectively. 
 
\begin{figure*}
 \begin{center}
\includegraphics[scale=0.25]{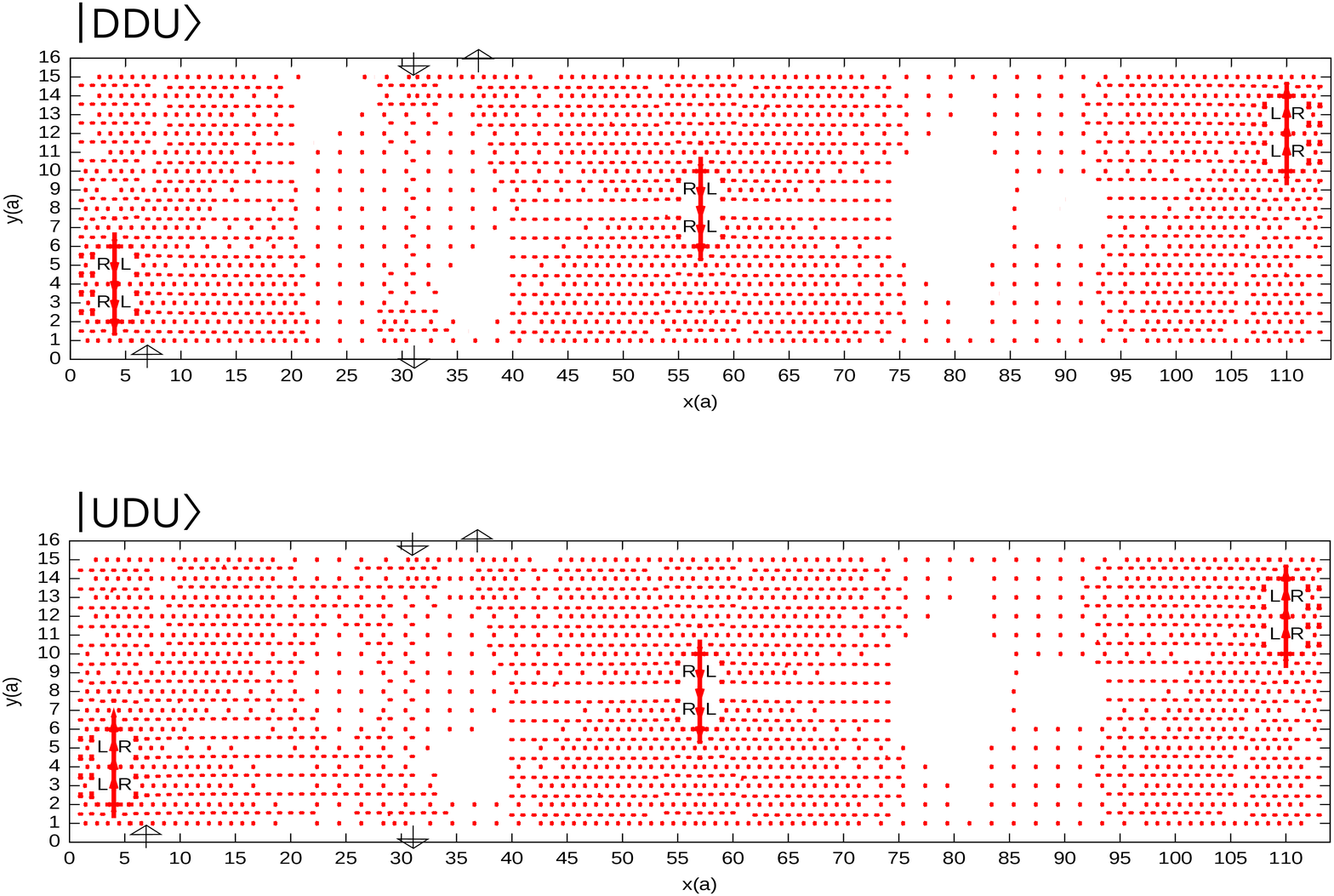}
\newline
 \end{center}
 \caption{
The current distribution of bottom two states of the three DCQ qubit system, in case external currents are fed at $(7,1)$ and $(31,15)$, drained at $(31,1)$ and $(37,15)$. . The two states are indicated as $\mid {\rm DDU} \rangle$ and $\mid {\rm UDU} \rangle$, respectively. }
 \label{3qubitDCQt2}
\end{figure*}

\begin{figure*}
 \begin{center}
\includegraphics[scale=0.25]{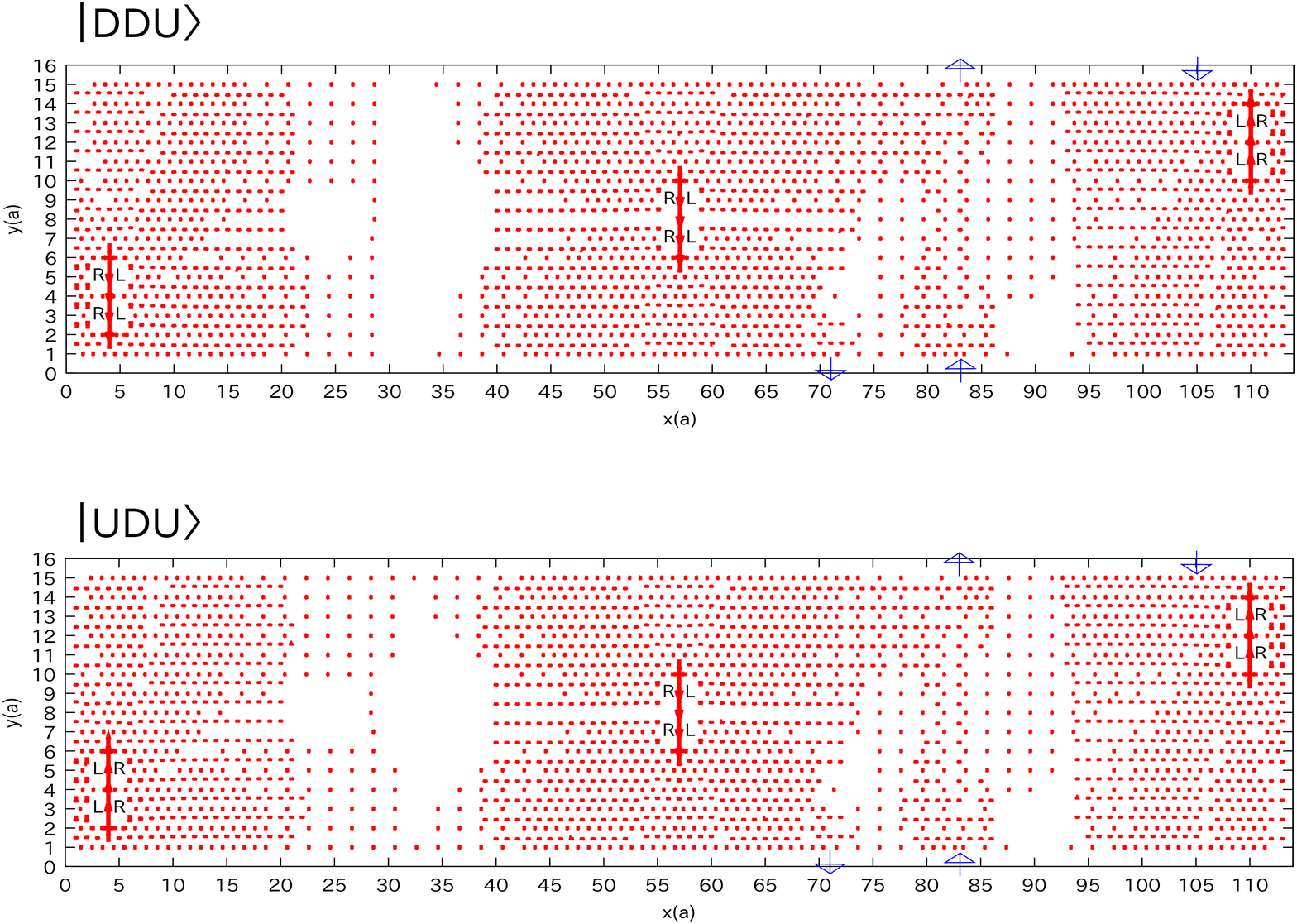}
\newline
 \end{center}
 \caption{
The current distribution of bottom two states of the three DCQ qubit system, in case external currents are fed at $(83,1)$ and $(105,15)$, drained at $(71,1)$ and $(83,15)$.  The two states are indicated as $\mid {\rm DDU} \rangle$ and $\mid {\rm UDU} \rangle$, respectively. }
 \label{3qubitDCQt3}
\end{figure*}

\begin{figure*}
 \begin{center}
\includegraphics[scale=0.25]{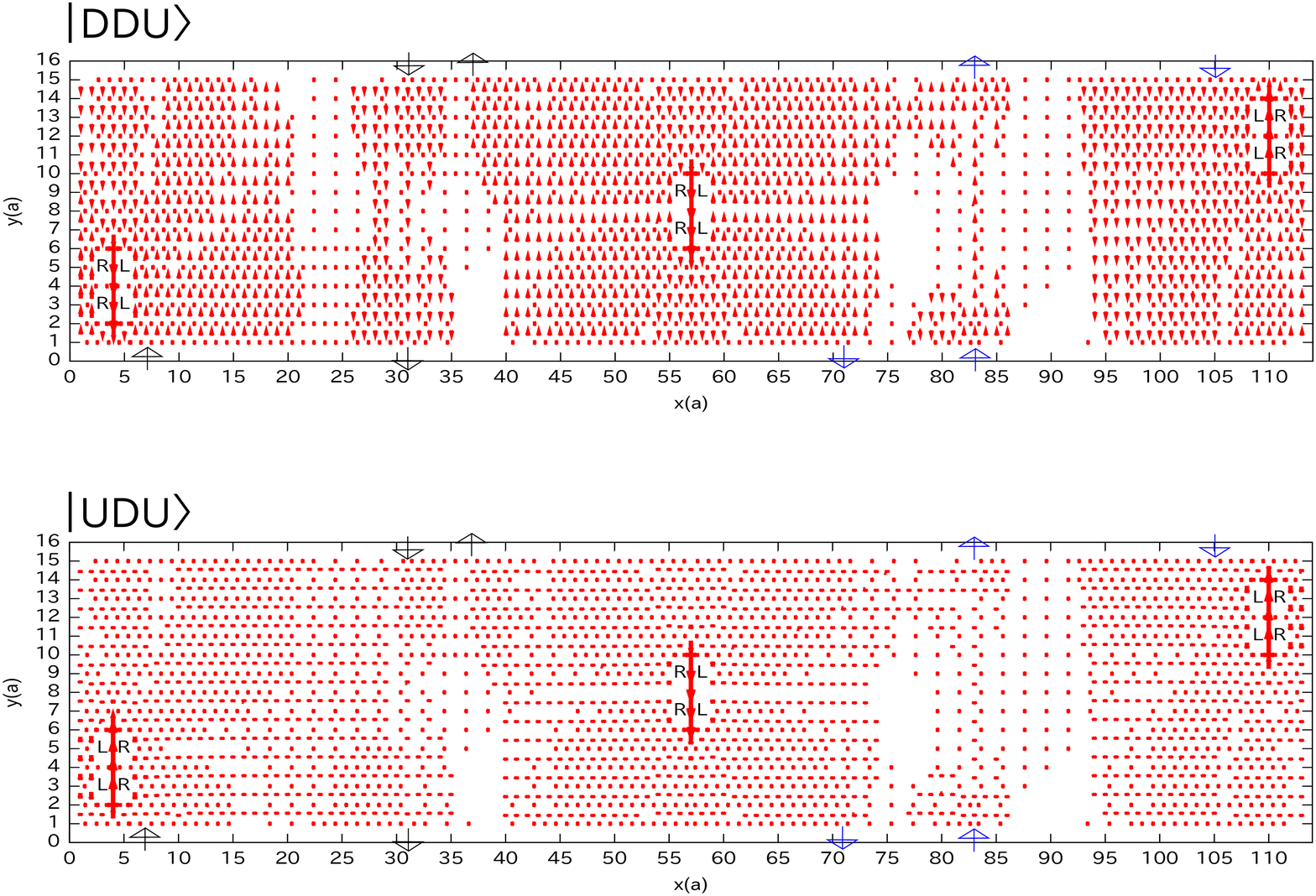}
\newline
 \end{center}
 \caption{
The current distribution of bottom two states of the three DCQ qubit system, in case external currents are fed at $(7,1)$, $(83,1)$,$(31,15)$ and $(105,15)$, drained at $(31,1)$,$(71,1)$,$(37,15)$ and $(83,15)$.  The two states are indicated as $\mid {\rm DDU} \rangle$ and $\mid {\rm UDU} \rangle$, respectively. }
 \label{3qubitDCQt4}
\end{figure*}

 In this section, we only feed external currents to couple two neighboring DCQs. For example, bottom two states in case ${\rm J}_{\rm ex}^{\rm 4} =0.02(2et/\hbar)$ to couple left two DCQs are shown in Fig.~\ref{3qubitDCQt2}. Bottom two states in case ${\rm J}_{\rm ex}^{\rm 5}=0.02(2et/\hbar)$ to couple right two DCQs are shown in Fig.~\ref{3qubitDCQt3}. Bottom two states in case ${\rm J}_{\rm ex}^{\rm 4}=0.04(2et/\hbar)$ and ${\rm J}_{\rm ex}^{\rm 5}=0.02(2et/\hbar)$ to couple left and right two DCQs are shown in Fig.~\ref{3qubitDCQt4}. 

Transition dipole moment of this system is shown in Table.~\ref{tdp3q}. $y$ component of transition dipole moment correspond to the transition of center DCQ is a little larger than that of side DCQs. This is because side DCQs are near the edge of CuO$_2$ plane compared to center DCQ. 

\begin{figure*}
 \begin{center}
\includegraphics[scale=0.45]{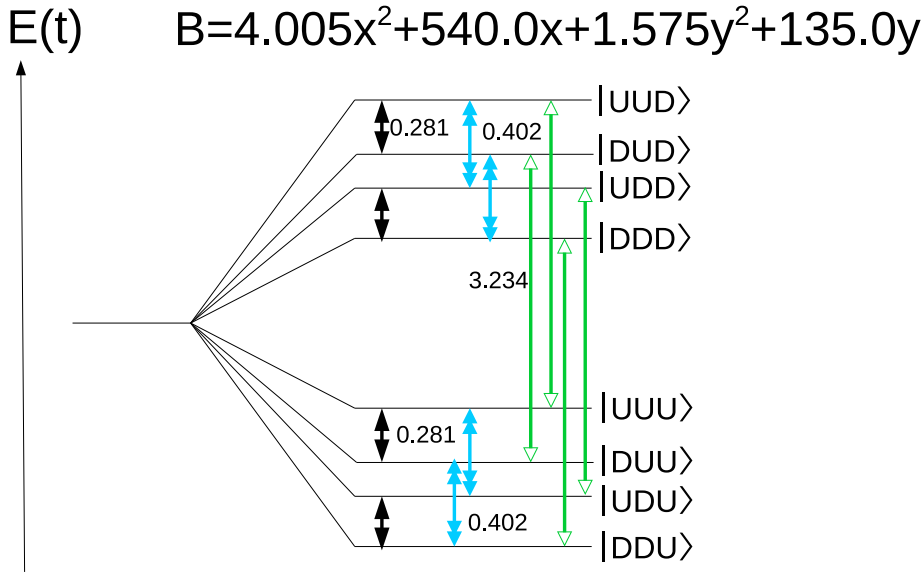}
\newline
 \end{center}
 \caption{
Energy levels of three DCQ qubit system. Left : $B=0$ and ${\rm J}_{\rm ex}=0$. Right : $B=4.005x^2+540.0x+1.575y^2+135.0y(T)$. The unit of x and y is $(a)$.}
 \label{3qiqene}
\end{figure*}

\begin{figure*}
 \begin{center}
\includegraphics[scale=0.25]{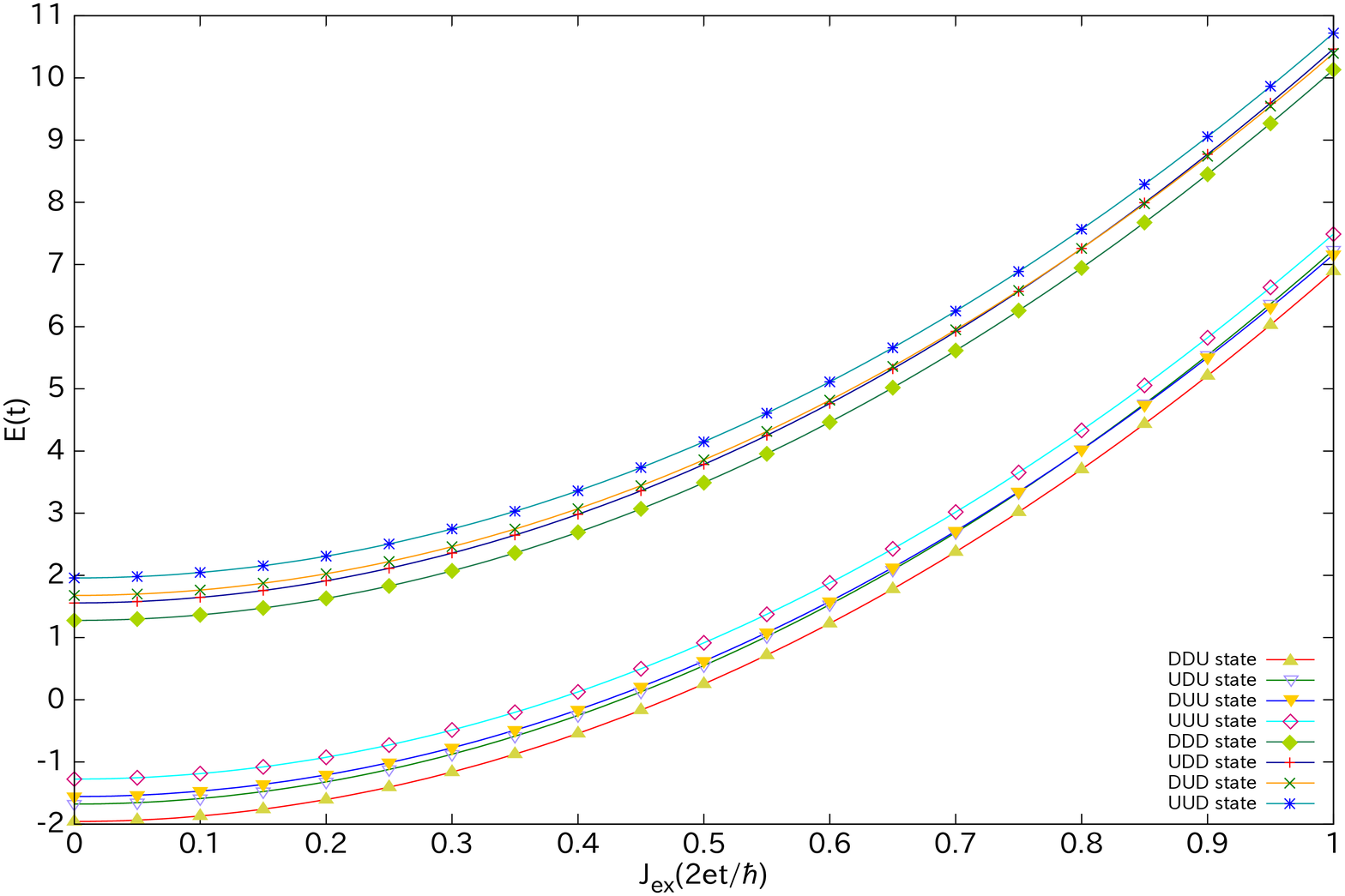}
\newline
 \end{center}
 \caption{Summed single particle energy of 8 states of three DCQ qubit system in case external currents are fed at $(7,1)$ and $(31,15)$ and drained at $(31,1)$ and $(37,15)$. When external current density is more than $0.7(2et/\hbar)$, $\mid {\rm UDU} \rangle$ states become above $ \mid {\rm DUU} \rangle$ state, and $\mid {\rm UDD} \rangle$ states become above $\mid {\rm DUD}\rangle$ state, respectively.
 }
 \label{enechanget2}
\end{figure*}

\begin{figure*}
 \begin{center}
\includegraphics[scale=0.25]{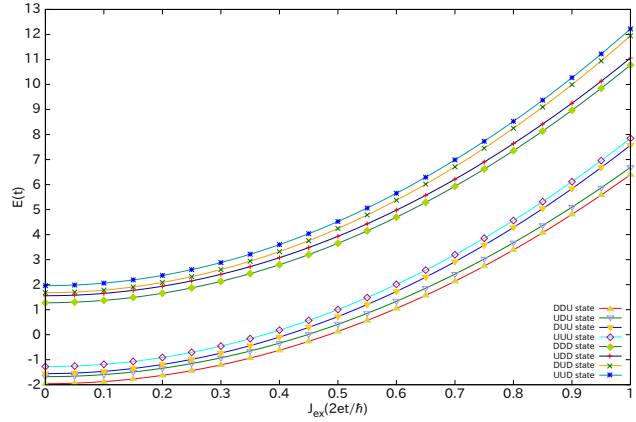}
\newline
 \end{center}
 \caption{Summed single particle energy of 8 states of three DCQ qubit system in case external currents are fed at $(83,1)$ and $(105,15)$ and drained at $(71,1)$ and $(83,15)$. 
 }
 \label{enechanget3}
\end{figure*}

\begin{figure*}
 \begin{center}
\includegraphics[scale=0.25]{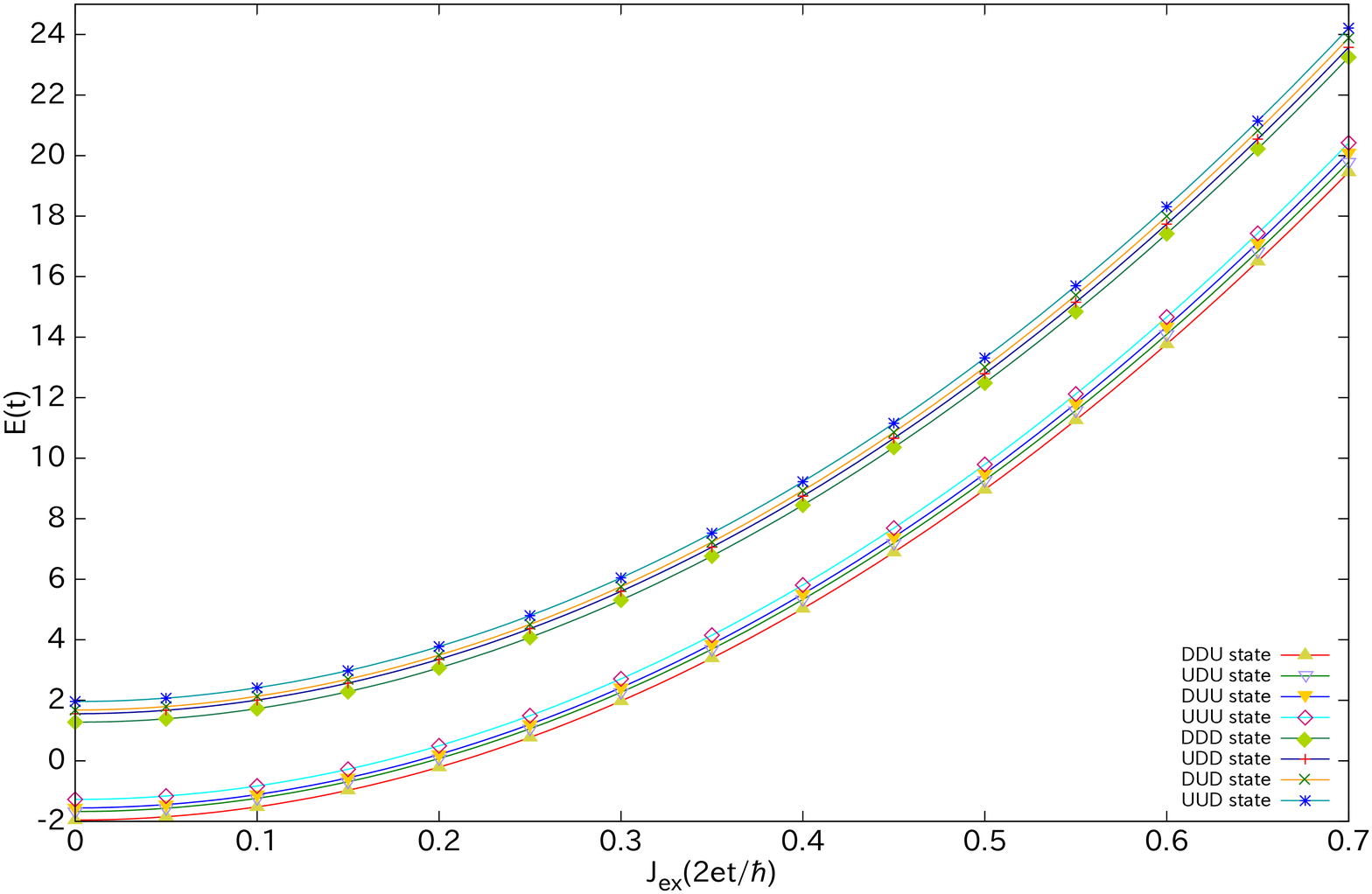}
\newline
 \end{center}
 \caption{Summed single particle energy of 8 states of three DCQ qubit system in case external currents are fed at $(7,1)$,$(83,1)$,$(31,15)$ and $(105,15)$ and drained at $(31,1)$,$(71,1)$,$(37,15)$ and $(83,15)$.  When external current density is more than $0.7(2et/\hbar)$, $\mid {\rm UDU} \rangle$ states become above $ \mid {\rm DUU} \rangle$ state, and $\mid {\rm UDD} \rangle$ states become above $\mid {\rm DUD}\rangle$ state, respectively.
 }
 \label{enechanget4}
\end{figure*}

Energy levels of three-qubit system are shown in Fig.~\ref{3qiqene}. They are split by inhomogenious magnetic field to remove all degeneracies of 8 states. Coupling of two DCQs occurs as the change of energy levels for the density of feeding external currents. Figs.~\ref{enechanget2},\ref{enechanget3},\ref{enechanget4} shows the change of energy levels v.s. feeding external currents to couple left two DCQs, right two DCQs and all DCQs, respectively. In case ${\rm J}_{\rm ex}^{\rm 4} \geq 0$ and left two DCQs are coupled, changing of energy levels depends on the states of coupled two DCQs. For example, $\mid DDU \rangle$ and $\mid DDD \rangle$ states have nearly same propotional coefficients.  Thus, left two DCQs are coupled. When external current density ${\rm J}_{\rm ex}^{\rm 4}$ is more than $0.7(2et/\hbar)$, $\mid {\rm UDU} \rangle$ states become above $ \mid {\rm DUU} \rangle$ state, and $\mid {\rm UDD} \rangle$ states become above $\mid {\rm DUD}\rangle$ state, respectively. In case ${\rm J}_{\rm ex}^{\rm 5} \geq 0$ and right two DCQs are coupled, changing of energy levels depends on the states of right two DCQs. In case  external currents satisfy ${\rm J}_{\rm ex}^{\rm 4}:{\rm J}_{\rm ex}^{\rm 5}=2:1$ and all DCQs are coupled, changing of energy levels is different for all states.

\begin{table}[h]
\caption{
Transition dipole moment of three DCQ qubit system. Elements in upper triangle region are $y$ component and elements in lower triangle region are $x$ component.
}\label{tdp3q}
\begin{center}
\begin{tabular}{c|c|c|c|c|c|c|c|c|c} \hline
 & \multicolumn{9}{|c}{$\mu^y(10^{-30}Cm)$} \\ \hline
&state&DDU&UDU&DUU&UUU&DDD&UDD&DUD&UUD \\ \cline{2-10}
&DDU&0&9.594&10.868&0&9.595&0&0&0 \\ \cline{2-10}
&UDU&0.061&0&0&10.868&0&9.595&0&0 \\ \cline{2-10}
&DUU&0&0&0&9.593&0&0&9.595&0 \\ \cline{2-10}
$\mu^x$&UUU&0&0&0.061&0&0&0&0&9.595 \\ \cline{2-10}
&DDD&0.061&0&0&0&0&9.593&10.868&0 \\ \cline{2-10}
&UDD&0&0.061&0&0&0.061&0&0&10.868 \\ \cline{2-10}
&DUD&0&0&0.061&0&0&0&0&9.594 \\ \cline{2-10}
&UUD&0&0&0&0.061&0&0&0.061&0 \\ \hline

\end{tabular}
\end{center}
\end{table}

\section{Splitting of energy levels by feeding external currents}\label{4}
We show the splitting character of three DCQ qubit system by feeding external currents here. We feed and drain external currents near the each DCQ qubits. We treat feeding external currents as classical purturbation that changes the energy in parabolic form of each feeding external current. The datas we sampled for two or more external feeding currents are sampled by one is varied and other are fixed. 

\begin{figure*}
 \begin{center}
\includegraphics[scale=0.25]{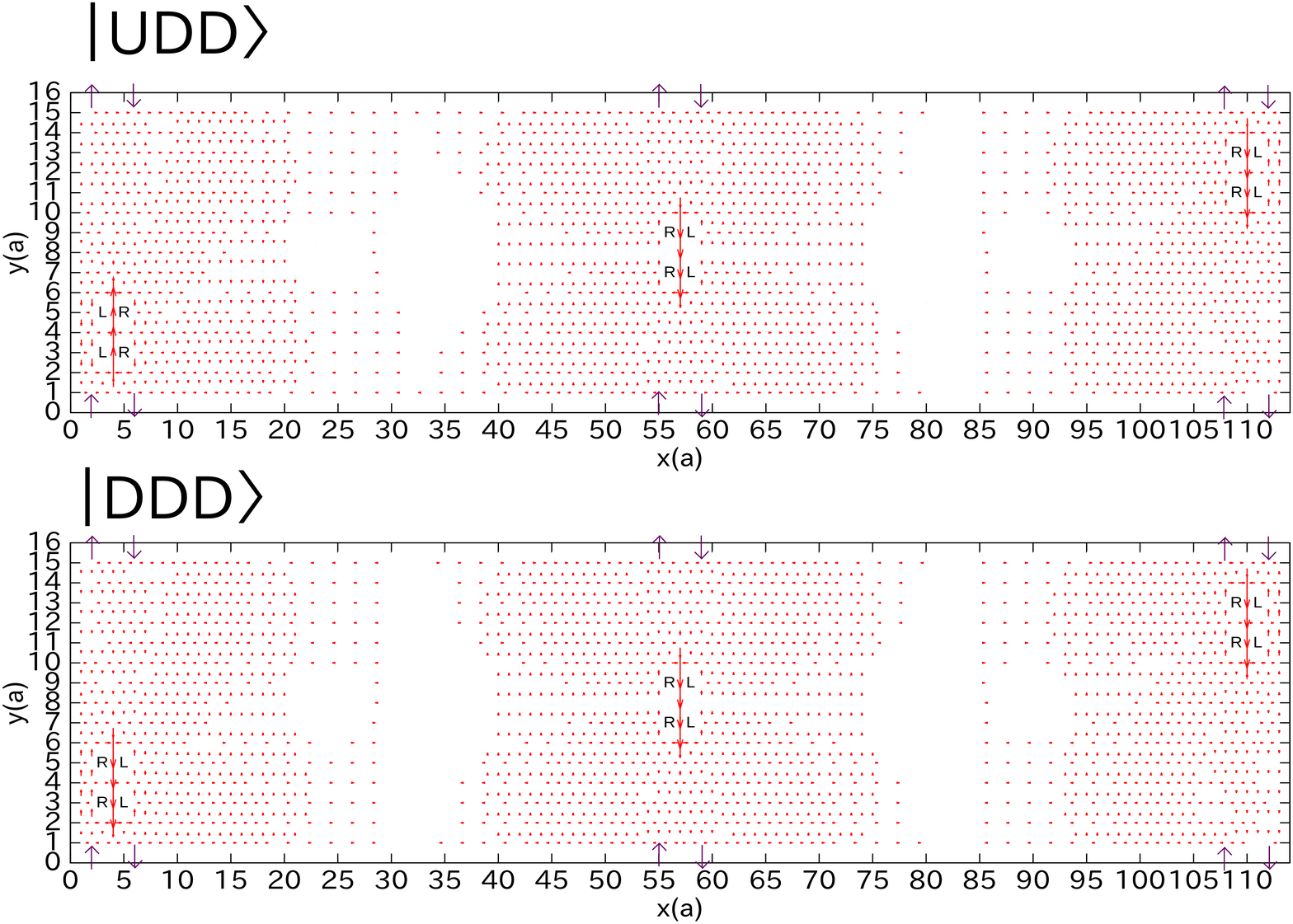}
\newline
 \end{center}
 \caption{Current distribution of two bottom states of 3 DCQ qubit system. Feeding external currents satisfy $J_{\rm ex}^{\rm 1}=0.002(2et/\hbar), J_{\rm ex}^{\rm 2}=0.004(2et/\hbar), J_{\rm ex}^{\rm 3}=0.008(2et/\hbar), J_{\rm ex}^{\rm 4}=J_{\rm ex}^{\rm 5}=0(2et/\hbar)$.
 }
 \label{DCQ3qiqsepex}
\end{figure*}

 We show the current distribution of three DCQ qubit system split by external currents in Fig.~\ref{DCQ3qiqsepex}. Their draining points and feeding points are same as those in Figs.~\ref{3qubitDCQt2},\ref{3qubitDCQt3},\ref{3qubitDCQt4}.

\begin{figure*}
 \begin{center}
\includegraphics[scale=0.25]{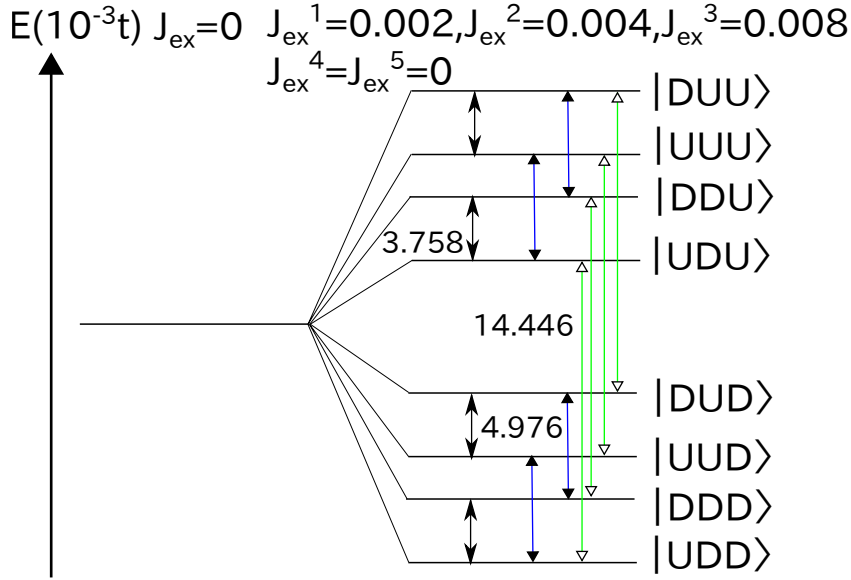}
\newline
 \end{center}
 \caption{Energy levels of 8 states of 3 DCQ qubit system. Left: $J_{\rm ex}^{\rm 1}=J_{\rm ex}^{\rm 2}=J_{\rm ex}^{\rm 3}=J_{\rm ex}^{\rm 4}=J_{\rm ex}^{\rm 5}=0$. Right: $J_{\rm ex}^{\rm 1}=0.002(2et/\hbar), J_{\rm ex}^{\rm 2}=0.004(2et/\hbar), J_{\rm ex}^{\rm 3}=0.008(2et/\hbar), J_{\rm ex}^{\rm 4}=J_{\rm ex}^{\rm 5}=0(2et/\hbar)$.
 }
 \label{enechangeexsep}
\end{figure*}

\begin{figure*}
 \begin{center}
\includegraphics[scale=0.25]{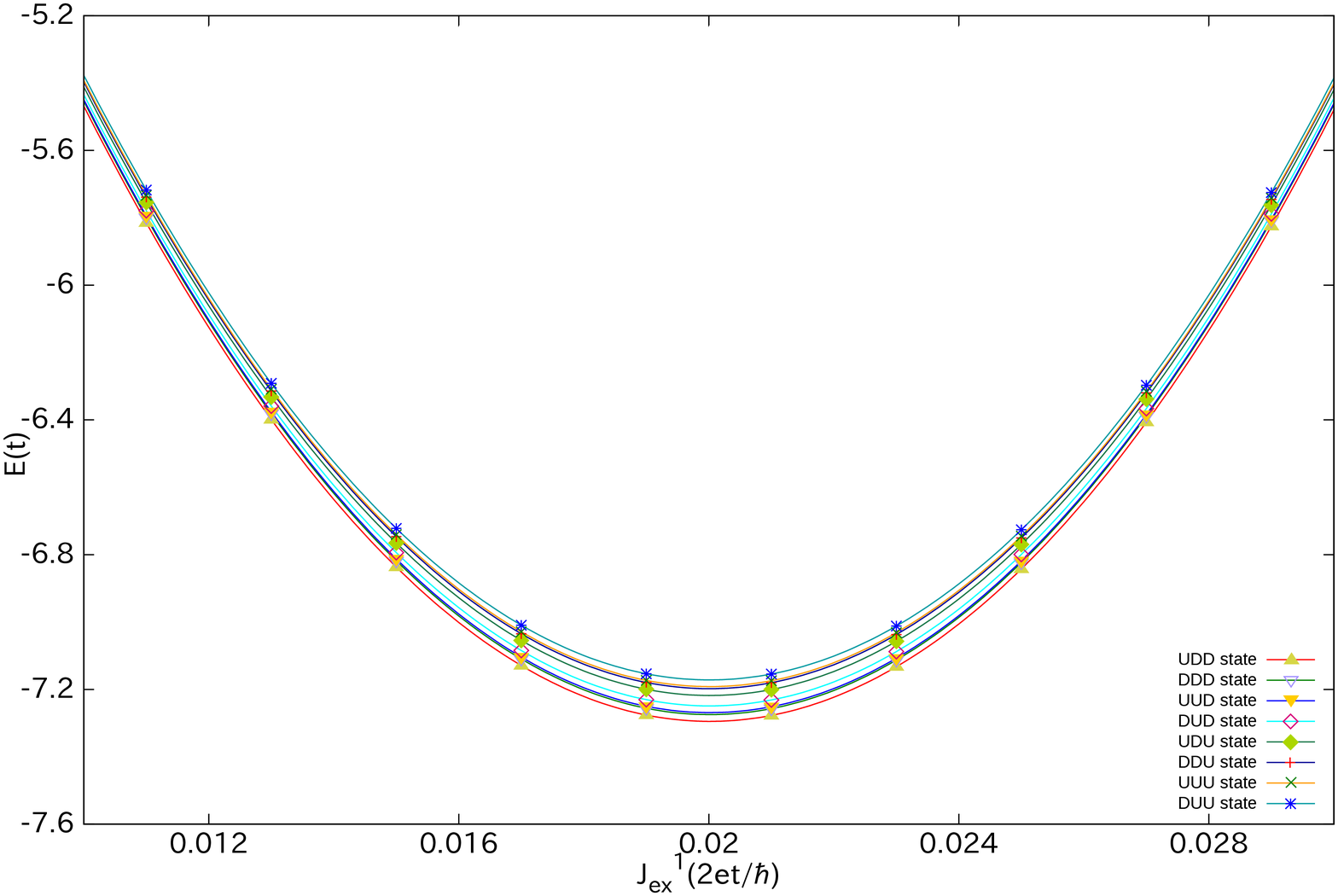}
\newline
 \end{center}
 \caption{Summed single particle energy of 8 states of three DCQ qubit system in case external currents satisfy $J_{\rm ex}^{\rm 1}:J_{\rm ex}^{\rm 2}:J_{\rm ex}^{\rm 3}=1:2:4$.
 }
 \label{enechangesepex}
\end{figure*}

 Energy levels in case that left DCQ is fed/drained $J_{\rm ex}^{\rm 1}=0.002(2et/\hbar)$, center DCQ is fed/drained $J_{\rm ex}^{\rm 2}=0.004(2et/\hbar)$, and right DCQ is fed/drained $J_{\rm ex}^{\rm 3}=0.008(2et/\hbar)$, respectively are shown in Fig.~\ref{enechangeexsep}. Fig.~\ref{enechangesepex} is the energy levels v.s. external current density $J_{\rm ex}^{\rm 1}$ that satisfies $J_{\rm ex}^{\rm 1}:J_{\rm ex}^{\rm 2}:J_{\rm ex}^{\rm 3}=1:2:4$. The minimums of energy levels are all same value $0.02(2et/\hbar)$.

 \begin{table}[h]
\begin{center}
\caption{
Transition dipole moment of three DCQ qubit system differenciated the environments of each qubit by feeding external currents. Elements in upper triangle region are $y$ component and elements in lower triangle region are $x$ component.
}
\begin{tabular}{c|c|c|c|c|c|c|c|c|c} \hline
 & \multicolumn{9}{|c}{$\mu^y(10^{-30}C\cdot m)$} \\ \hline
& state&UDD&DDD&UUD&DUD&UDU&DDU&UUU&DUU \\ \cline{2-10}
&UDD&&9.594&10.869&0&9.594&0&0&0 \\ \cline{2-10}
&DDD&0.061&&0&10.869&0&9.594&0&0 \\ \cline{2-10}
&UUD&0&0&&9.594&0&0&9.594&0 \\ \cline{2-10}
$\mu^x$&DUD&0&0&0.061&&0&0&0&9.594 \\ \cline{2-10}
&UDU&0.061&0&0&0&&9.594&10.869&0 \\ \cline{2-10}
&DDU&0&0.061&0&0&0.061&&0&10.869 \\ \cline{2-10}
&UUU&0&0&0.061&0&0&0&&9.594 \\ \cline{2-10}
&DUU&0&0&0&0.061&0&0&0.061& \\ \hline

\end{tabular}
\end{center}
\end{table}

\begin{figure*}
 \begin{center}
\includegraphics[scale=0.25]{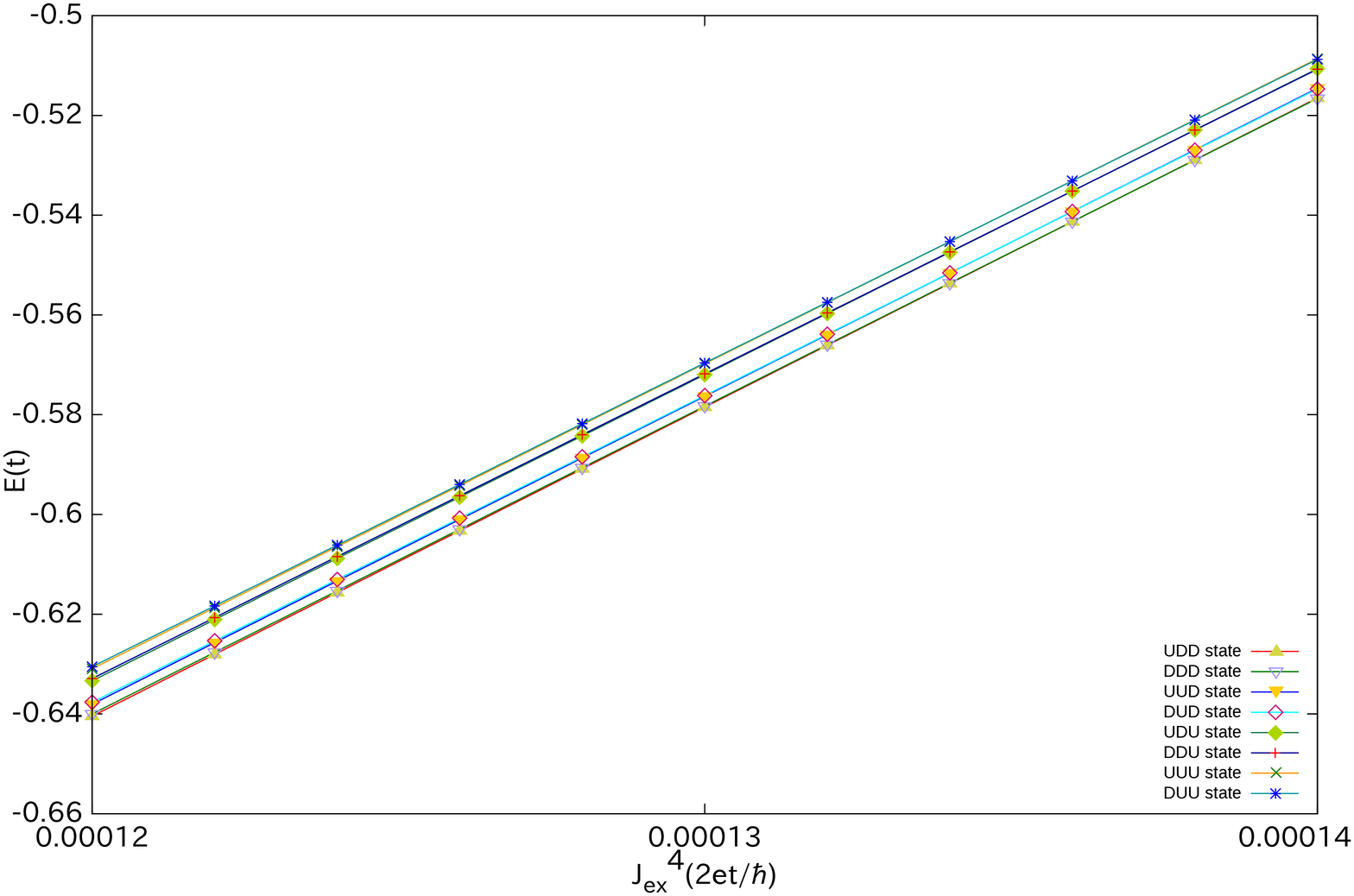}
\newline
 \end{center}
 \caption{Summed single particle energy of 8 states of three DCQ qubit system v.s. ${\rm J}_{\rm ex}^{\rm 4}$ in case external currents satisfy $J_{\rm ex}^{\rm 1}=0.002(2et/\hbar), J_{\rm ex}^{\rm 2}=0.004(2et/\hbar),J_{\rm ex}^{\rm 3}=0.008(2et/\hbar)$ and ${\rm J}_{\rm ex}^{\rm 5}=0(2et/\hbar)$.
 }
 \label{enechangesepext2}
\end{figure*}

\begin{figure*}
 \begin{center}
\includegraphics[scale=0.25]{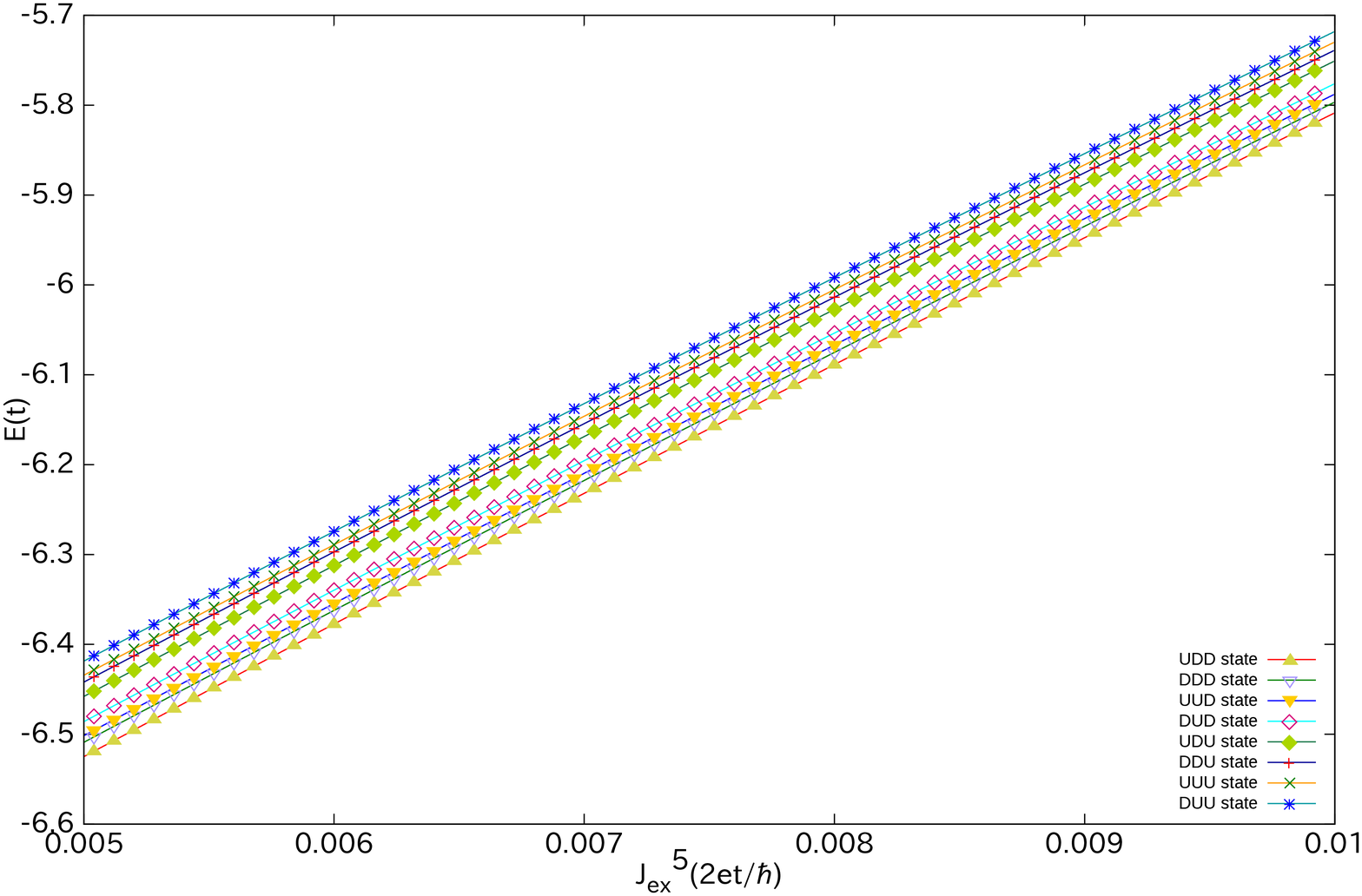}
\newline
 \end{center}
 \caption{Summed single particle energy of 8 states of three DCQ qubit system v.s. ${\rm J}_{\rm ex}^{\rm 5}$ in case external currents satisfy $J_{\rm ex}^{\rm 1}=0.002(2et/\hbar), J_{\rm ex}^{\rm 2}=0.004(2et/\hbar),J_{\rm ex}^{\rm 3}=0.008(2et/\hbar)$ and ${\rm J}_{\rm ex}^{\rm 4}=0(2et/\hbar)$.
 }
 \label{enechangesepext3}
\end{figure*}

\begin{figure*}
 \begin{center}
\includegraphics[scale=0.25]{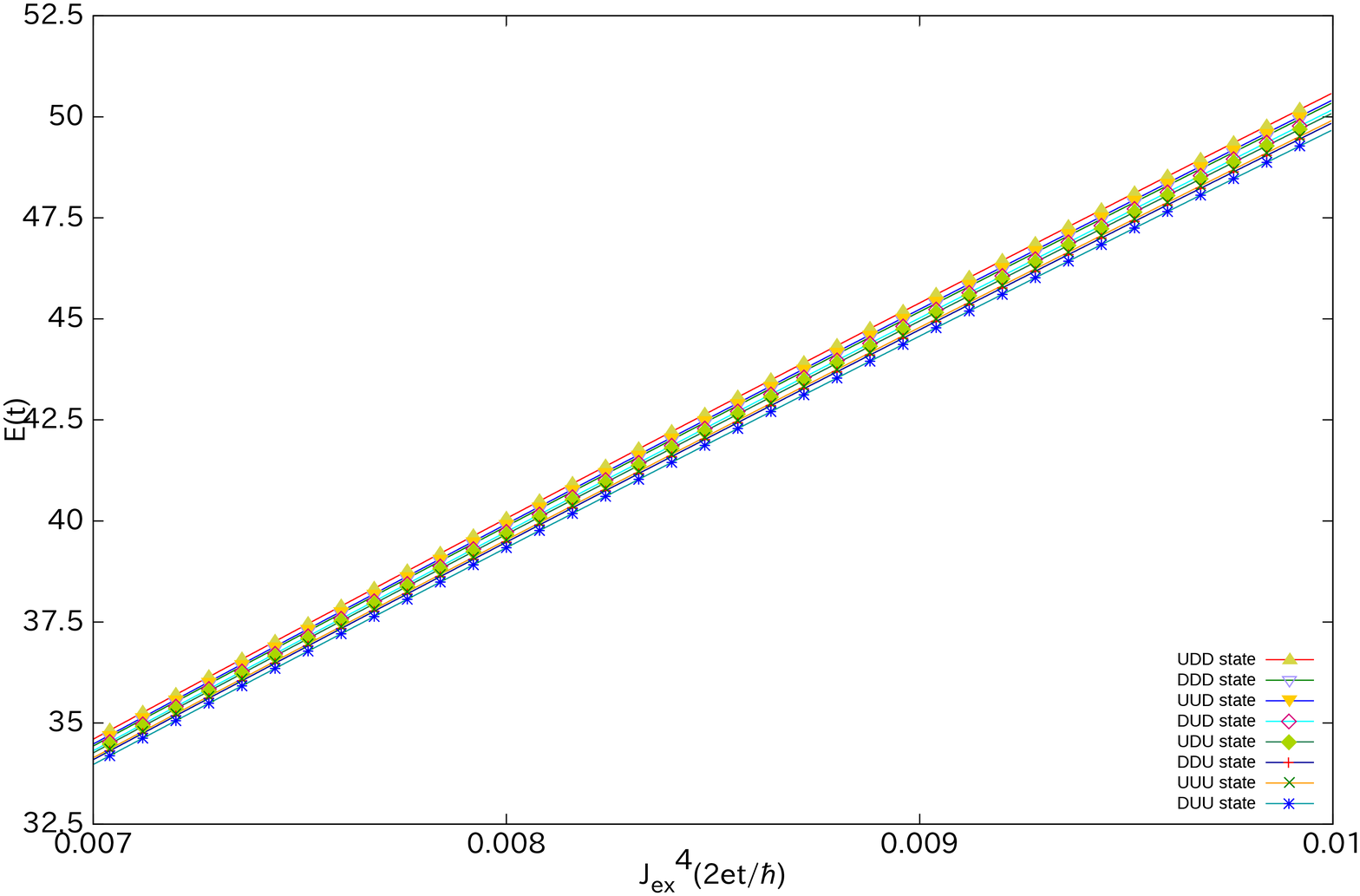}
\newline
 \end{center}
 \caption{Summed single particle energy of 8 states of three DCQ qubit system v.s. ${\rm J}_{\rm ex}^{\rm 4}$ in case external currents satisfy $J_{\rm ex}^{\rm 1}=0.002(2et/\hbar), J_{\rm ex}^{\rm 2}=0.004(2et/\hbar),J_{\rm ex}^{\rm 3}=0.008(2et/\hbar)$ and ${\rm J}_{\rm ex}^{\rm 4}:{\rm J}_{\rm ex}^{\rm 5}=1:2$.
 }
 \label{enechangesepext4}
\end{figure*}

Fig.~\ref{enechangesepext2} shows the energy levels v.s. ${\rm J}_{\rm ex}^{\rm 4}$. In that case,  feeding external currents satisfy $J_{\rm ex}^{\rm 1}=0.002(2et/\hbar), J_{\rm ex}^{\rm 2}=0.004(2et/\hbar), J_{\rm ex}^{\rm 3}=0.008(2et/\hbar)$ and ${\rm J}_{\rm ex}^{\rm 5}$ is zero. Splitting of energy levels and coupling of left and center DCQs are confirmed to be occur together.
Fig.~\ref{enechangesepext3} shows the energy levels v.s. ${\rm J}_{\rm ex}^{\rm 5}$. In that case,  feeding external currents satisfy $J_{\rm ex}^{\rm 1}=0.02(2et/\hbar), J_{\rm ex}^{\rm 2}=0.04(2et/\hbar), J_{\rm ex}^{\rm 3}=0.08(2et/\hbar)$ and ${\rm J}_{\rm ex}^{\rm 5}$ is zero. The coupling of qubits coincide with splitting of energy levels sometimes requires changing the values of feeding external currents for splitting because the magnitude of coupling is limited by the magnitude of feeding external current for splitting. Fig.~\ref{enechangesepext4} shows the energy levels v.s. ${\rm J}_{\rm ex}^{\rm 4}$. In that case,  feeding external currents satisfy $J_{\rm ex}^{\rm 1}=0.02(2et/\hbar), J_{\rm ex}^{\rm 2}=0.04(2et/\hbar), J_{\rm ex}^{\rm 3}=0.08(2et/\hbar)$ and ${\rm J}_{\rm ex}^{\rm 4}:{\rm J}_{\rm ex}^{\rm 5}=1:2$. The coupling between left and center DCQs is far stronger than that of center and right DCQs.
Although, the intensity of coupling between center and right DCQs is far weaker than that of left and center DCQs. This is supposed that it is because draining point of $J_{\rm ex}^{\rm 5}$ is too close to center DCQ, thus, current flow between $(x, y)=(83, 1)$ and $(71, 1)$, $(83, 15)$ and $(105, 15)$ have small current density. Therefore, current distribution range of overlap current between center and left DCQs is so small that it can't couple them.

\newpage

\section{Concluding Remarks}\label{6}
In this work, it is confirmed that splitting of energy levels by feeding external currents coincide with coupling of qubits by them can be realized. This means that feeding external currents can be used also for splitting energy levels. Splitting all energy levels of qubit states by static magnetic fields will be more difficult as the number of qubit increases because the intensity of magnetic field realized constantly is limited. Hence, splitting energy levels by magnetic field will be difficult for many qubit. Though, coupling of any neighboring qubits requires some tequnics to decide the locations of source and drain points, SVILCs never require static magnetic fields for any quantum coupling and differentiation. It means that, SVILC qubits have no limit of downscaling due to the size of generators of static magnetic field, that both Ion-Trap and typical superconducting qubits inevitably have, thus, nano-sized differentiators of SVILC qubits can be realized. Magnetic fields only appear in the laser for quantum operation and readout process that SVILCs themselves emerge. Using feeding external currents as nano-sized couplers and differentiaters for nano-sized SVILC qubits, 100 physical qubit system of SVILC qubits in 2D-layer may be realized in a hundred of nanometer scale. Therefore, the advantage of SVILC qubits 1) and 2) are partially confirmed.  Next problem is to realize the qubit operation by crossing energy levels and quantum error correction. Arbitrary coupling is also remaining. When these probrems are solved all, the first portable quantum computers may be bringed in the world.
3D-layer of CuO$_2$ planes can be used as topological quantum computers using the fluxes made by SVILCs because the energy levels of all planes can be same by external currents too. 

\bibliographystyle{apsrev4-2}
\bibliography{tesbib}
\end{document}